# Time-correlation Transduction in Strong-field Quantum Electrodynamics

*Zairui Li[1*], Wesley Sims[1], Mirali Seyed Shariatdoust[3,6], Gabriel Howell[1], Thomas A. Searles[2], Sergio Carbajo[3-6]*

[1] *Department of Physics, Morehouse College, Atlanta, USA*
[2] *Electrical and Computer Engineering Department, University of Illinois Chicago, Chicago, USA*
[3] *Department of Electrical and Computer Engineering, University of California Los Angeles, Los Angeles, CA 90095, USA*
[4] *Department of Physics and Astronomy, University of California Los Angeles, Los Angeles, CA 90095, USA*
[5] *SLAC National Accelerator Laboratory, Stanford University, 2575 Sand Hill Road, Menlo Park, CA, USA*
[6] *California NanoSystems Institute, 570 Westwood Plaza, Los Angeles, CA 90095, USA*
*corresponding author: Zairui.li@morehouse.edu*

**Abstract**

Recent developments in high-power ultrafast optical technology and emerging theoretical frameworks in strong-field quantum electrodynamics (SF-QED) are unveiling nuanced differentiations between the semi-classical and full quantum mechanical descriptions of physical systems. Here we present a computational investigation of a novel technique for attosecond optical sensing through time correlation transduction (TCT) by investigating high-harmonic generation (HHG) as a representative SF-QED process. TCT is an experimental method to capture photon-electron interactions at higher harmonic orders by temporarily correlating the emitted and driving photon fields. This approach enables resolving the dynamical behavior of optically-driven strong-field phenomena in quantum materials such as Two-dimensional Materials and Dirac Semimetals down to 10 attosecond temporal resolution to discover a full quantum explanation. Predicting and measuring the transition between perturbative and non-perturbative regimes with attosecond resolution can deepen the understanding of SF-QED such as HHG. As such, we find that TCT is a powerful method to pave the way toward the functional characterization of quantum matter.

**I. Introduction**

As a result of recent developments in ultrafast optical technology, nuclear and electron dynamics can be studied at the attosecond time scale[3,4] using light sources such as extreme ultraviolet (XUV) or soft x-ray radiation[1,2] that are upconverting fundamental frequencies by high-harmonic generation (HHG). In an attosecond time frame, the energy gain of an electron accelerated over one Compton wavelength is a nonlinear process that requires a critical electric field. The process of Probing critical electric field level challenges quantum electrodynamics (QED) to describe a nonperturbative regime that is not fully covered by the standard time-dependent Schrödinger equation model[5,6]. In strong-field quantum electrodynamics (SF-QED), the motion of electrons on the atomic scale is hidden from direct experimental access. Nonperturbative QED and its transition from perturbative QED above the Schwinger critical field can be characterized with ultrafast and ultrahigh-intensity laser technology.

Importantly, vibrant nonlinear phenomena such as rapid acceleration of electrons and ions[7], copious electron-positron pair production[8], and exotic light-by-light scattering[9] can be explained within SF-QED frameworks.

Under strong-field conditions, the characterizations of HHG rely on spectral analysis and the signature of energy exchange between electrons and photons[10,11] according to the three-step model. Firstly, an electron tunnels out from the atomic potential suppressed by the intense driving field (i), then accelerated in the continuum by the driving field (ii), and eventually under certain conditions can return to the ion and recombine (iii), emitting a high energy photon. This sequence repeats over time, producing a spectrum of harmonics as comb emission. Classical approaches to this model overlook the interaction between the electron and its core, leading to discrepancies when compared with experimental data from high harmonic spectroscopy[12]. To address this variation, the strong-field approximation (SFA) theory adjusts by acknowledging the role of the positively charged ionic core and its interaction with the electron post-ionization, suggesting a short-range core potential. Further disagreement is revealed via the outcome of the attoclock experiment, and the SFA alone proves inadequate[13]. The inclusion of Coulomb corrections[14], integrating the long-distance interactions between the departing electron and its parent ion, refines the induced dipole representation in HHG. This adjustment, merging the electron's bound state wave function driven by the laser field, yields a closer match to experimental observations. Recent advancements in techniques, such as the attosecond streak camera[15], the attoclock[16], attosecond transient absorption[17], RABBIT[18], and high harmonic spectroscopy[18], have enabled precise measurements of ionization times to an accuracy of tens of attoseconds, across both single-photon and multi-photon processes. These achievements underscore the importance of a robust theoretical foundation to accurately interpret experimental outcomes because establishing a coherent framework that integrates classical notions of time with quantum wavefunctions and experimental data is essential. The timing measurements are derivations from these experiments and are highly dependent on the methodology employed, highlighting the intricate relationship between theory and practice in the field.

Recent advancements propose a formalism for HHG that incorporates a full-quantum description[19]. Unlike the above discussion of the semi-classical perspective, a full quantum correction not only considers the driving field and electron states within the quantum regime but also models the quantization of emitted radiation. Consider that electron states are provided by a strong laser field with multiple quantized field modes, and all the modes are initially in a ground state[20]. This quantum perspective regards HHG as a spontaneous emission from these time-dependent states. Hence, photon generation occurs in unoccupied modes and accounts for the nonlinear nature of frequency conversion through transitions between laser-dressed electronic states. This approach introduces a breakdown of the dipole approximation to the emitted photons and suggests each HHG photon carries all frequencies of the comb with attosecond timescales. By providing a framework that closely aligns with quantum electrodynamical theories, a full quantum explanation addresses the limitations of semi-classical models; although it is limited in scope and requires further experimental development.

In this work, we introduce a technique of ultrafast time-correlation transduction (TCT) to compare semi-classical and full quantum system behaviors from experimental practice. This presents an opportunity to explore the distinctions between nonperturbative[21] and perturbative[22] processes, highlighting their roles in all the HHG three steps within strong field environments. The strategy is to compare both semi-classical system and full quantum system behaviors. In the semi-classical perspective, a perturbative system is associated with individual frequencies in a mixed state. Conversely, nonperturbative systems imply each photon exists in a quantum superposition encompassing all frequencies within the comb. Advancing to a fully quantum domain, harmonic emissions become immutable by external weak field probing as perturbations. TCT provides attosecond-level measurements to observe the quantum regime's onset. This method is capable of cross-correlating wave packets of different natures and energy (e.g. bosons and fermions across the spectrum) at operating GHz data rate levels. TCT precisely identifies the emission timings of wave packets using the simulation outcome of HHG temporal dynamics in solids[23–25]. The results demonstrate the accuracy of TCT in discerning the dynamics of the three-step HHG process, potentially deepening our understanding of quantum photo-electrodynamics by providing evidence of higher harmonic order entanglement suggested by SF-QED. Beyond HHG, attosecond-level TCT analysis opens avenues for investigating other SF-QED physics, including the study of axion-like Particles (ALPs). These particles, potentially elucidated through time-resolved Kerr rotation due to their polarization-sensitive coupling[26], offer explanations for astrophysical phenomena such as the universe's transparency to high-energy gamma rays[27,28] and the mysterious 3.55 keV emission line observed in galaxy clusters[29,30]. Investigating these phenomena promises to enhance our comprehension of HHG and its implications for attosecond science, quantum optics, QED, and quantum information science.

**II. Methods**

Figure 1a depicts the proposed experimental setup of TCT. In the initial stage, a high energy laser pulse ~120 μJ energy and short duration (~120 fs) is divided into two orthogonal wave packets with 1550 nm wavelength via a beam splitter (BS1). Referring to the upper optical path, one wave packet is down-converted to a longer wavelength by an optical parametric amplifier (OPA), tuning it to the fundamental frequency (ω) of samples, which falls within a wavelength range of 3.8 μm to 4.8 μm. The sample is positioned between a pair of Yttrium orthovanadate ($YVO_4$) polarizers to isolate the polarization state. As Figure 1a demonstrates, ω′ represents the fundamental frequency of a specific material. 3ω′ and 5ω′ represent 3rd and 5th higher-order emissions with an induced time delay. Subsequently, an additional linear polarizer (P1) is used to retain the *signal*'s orthogonality. Lastly, a harmonic selector (band-pass filter BPF) isolates the harmonic order which has a wavelength near 1550 nm as a signal before a balanced optical cross-correlator (BOC). At the lower path, the *reference* wave packet propagates through an optical path delay (OPD) to control the temporal synchronization with the pump. After recombining the *signal* and *reference via* BS2, the post-HHG temporal delay encoded into the *signal* disrupts the initial synchronization between the *reference* and pump. This alteration in time correlation is measurable by BOC-based correlation[31].

Before the BOC, the *signal* and *reference* are orthogonal for the nonlinear up-conversion process. After propagating through a nonlinear crystal (e.g. PPKPD), the first sum-frequency wave packet is generated due to the overlap region and its amplitude is proportional to their temporal overlap. Under the same principle, by reflecting the *signal-reference* pair back through the nonlinear crystal, the second sum-frequency wave packet is generated. Because of the dispersion due to the initial propagation through the nonlinear crystal, the temporal overlap changes resulting in an amplitude difference ΔV. This double-pass process is determined by a balanced optical detection system that results in a linear relation between ΔV and the time delay between *signal* and *reference* ΔT before the BOC. ΔT contains two types of delays: controllable delay and HHG-induced delay. The controllable delay $\Delta t_i$ is as a result of the OPD and the initial driving field's temporal configuration which can be set to $\Delta t_i = 0$. The *signal* that is encoded with ultrafast physical phenomena carries the other type of delay. In this study it refers to the time differences $\Delta t_e$ due to higher-order emissions. Thus,

$$\Delta T = \Delta t_i + \Delta t_e. \qquad (1).$$

To induce nonperturbative and perturbative regime changes, a weak secondary pump (WSP) corresponding to signal wavelength (~1550 nm as an example) is introduced externally as the green pulse demonstrated in Figure 1b. The WSP interacts with the excited area of the samplel during HHG at ~10 fs away from the original generation of the targeted harmonic order (at 1550 nm). The time configuration of WSP overlaps with the excitation duration of ~100 fs and maximizes the time delay of stimulated emission for higher response. In the perturbative regime, radiative recombination is induced by electron-core dipoles; such an induction can also be contributed by the WSP as a stimulation process. Figure 1b demonstrates a perturbative regime that introduces an additional time delay $\Delta t_{3rd}$ via WSP-stimulated emission at 3rd harmonic frequency. Thus, $\Delta T$ becomes, $\Delta T = \Delta t_i + \Delta t_e + \Delta t_{3rd}$. In contrast, SF-QED suggests a nonperturbative regime, in which electron transitions occur between quantized driving field states and quantized higher energy emission states. WSP now neither stimulates a transition nor provides additional vacuum states (ground states) due to its weak interaction. The secondary pumping process in the consideration of SF-QED provides a $\Delta t_{3rd} = 0$ or $\Delta t_{3rd} \neq 0$ for quantum mechanical description or semi-classical perspective. Since $\Delta t_i$ is tunable, $\Delta t_i + \Delta t_e$ can be set to zero results $\Delta T$ depend on $\Delta t_{3rd}$ only. In practice, the amplitude of $\Delta T$ can be reflected as the BOC's temporal sensitivity. In theory, as the contribution of the stimulated process $\Delta t_{3rd} > 0$ increases the sensitivity also increases. Conversely, for the entirety of the full quantum process $\Delta t_{3rd} = 0$, the sensitivity is deactivated. Therefore, TCT is capable of monitoring this dynamic change.

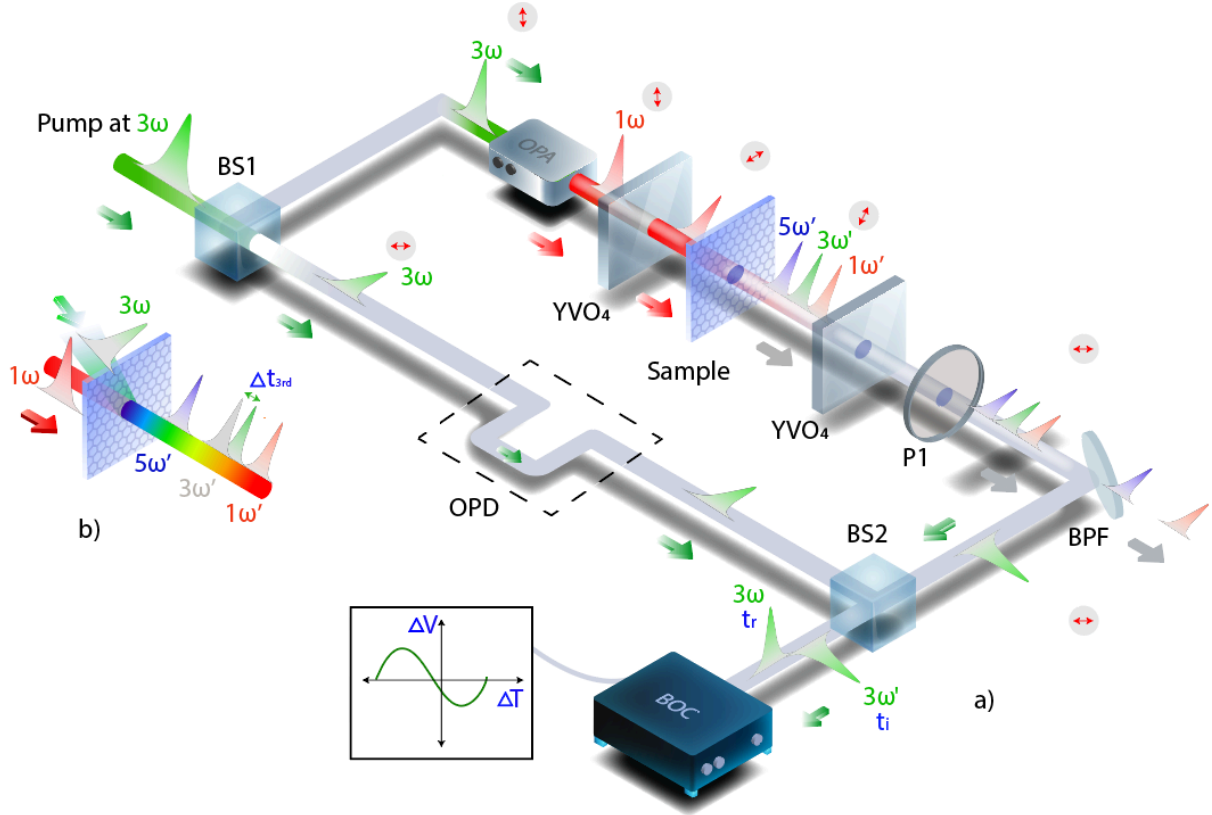

Figure 1: a) TCT for pump-probe time correlation measurement of HHG in quantum materials; BS1 and BS2 (beam splitters), OPD (optical path delay), YVO$_4$ (waveplate), P1 (polarizer), BPF (band pass filter) and BOC (balanced optical cross-correlator); b) a secondary weak pump stimulation, which induces a $\Delta t'$ delay for the 3rd order harmonic generation.

### III. Results

Using the testbed system described above, we present a range of simulations to evaluate the efficacy and robustness of the TCT in HHG from materials such as ZnO[23], MoS$_2$[32], and Cd$_3$As$_2$[25,24] and single-layer graphene[33] and including variations in noise levels and pulse shapes. As an example, the TCT output is depicted by blue to red curves in Figure 2a, which illustrates the BOC output of the 3rd HHG of ZnO measured in µV over a time delay range from - 1.5 to 1.5 ns. As a consequence of the material, an increase in fundamental bandgap step differences can be seen between the probe and reference pulses in the variation in wavelength from 1550 nm by increments of 15 nm. Under the consideration of a balanced photodetector with a responsivity of 0.88 A/W and an HHG quantum efficiency at a scale of -8e[34], the output at central wavelength 1550 nm at nearly balanced conditions, the BOC to attain a maximum response range of 150 µV within 1.5 ns. Multi-plots quantify the decreased values $\Delta V$ in BOC response due to deviation from the center wavelength. $\Delta V$ is linearly related to $\Delta T$ due to the transduction process. BOC detection provides time jitters on the order of 10 attoseconds[31] and a noise floor of nV/fs. Figure 1b is a zoomed-in of Figure 1a as highlighted, the linear region near zero-crossing. The linear region defines TCT's corresponding maximum sensitivity level of 13.6 pV/attosecond.

Reference to experimental time-frequency analysis[23–25], such sensitivity level is required to features temporal different of higher hommonic orders. Furthermore, simulated TCT sensitivities of $Cd_3As_2$, $MoS_2$ crystal, and mono-layer graphene as candidate materials within the working spectrum range are demonstrated in Figure 3.The sensitivity of a material in TCT and HHG experiments is influenced by its electronic structure, nonlinear optical properties, carrier dynamics, dimensionality, crystal symmetry, and external experimental conditions. Understanding these factors allows to select appropriate materials for higher probabilities to observe ultra fast phenomena. Each demonstrates a simulated BOC response (in W) for the materials with a delay range of -10 fs to 10 fs. And the corresponding near balanced outcome is summarized in Table I. The 3rd harmonic order emission of ZnO demonstrates the highest sensitivity due to the most significant deviation to central wavelength 1.55 $\mu$m. The last two columns of Table I introduce the time delay at zero crossing where $\Delta t_i + \Delta t_e = 0$ and the maximum sensitivity value corresponding to each material—these default values of TCT simulation as floor values for comparison with the experimental values.

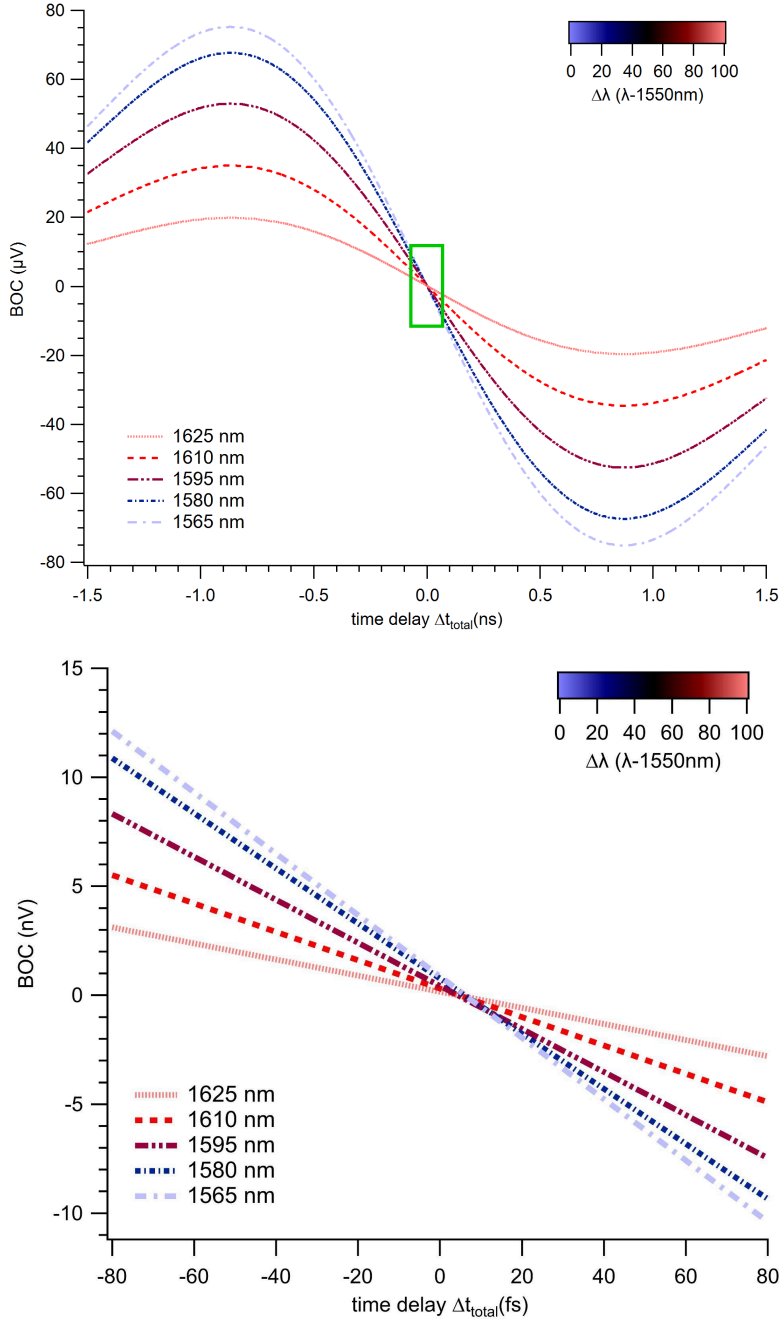

Figure 2: a (left) plots demonstrate the simulation results of TCT output ($\mu$V) concerning the accumulated time delay $\Delta T$ (ns). Multiple plots from blue to red that represent the signal wavelength from 1550 nm to 1700 nm in intervals of 50 nm. b (right) demonstrates linearity at zero-crossing of part a) ranging from -30 fs to 30 fs.

To define the sensitivity of the secondary probing method, at first two boundaries are described as $\Delta V_{max}(\Delta t_{3rd} = 10\,fs)$ and $\Delta V_{min}(\Delta t_{3rd} = 0)$. BOC output becomes $\Delta V = c_c \Delta V_{max} + c_q \Delta V_{min}$, where

$c_c$ represents the amplitude coefficient of WSP altered emission, $c_q$ represents the amplitude coefficient of emission that is nonperturbative. The sensitivity refers to $\frac{c_c}{c_c+c_q}$ because only $c_c$ affects BOC output. In addition, these amplitude coefficients are also related to pumping strength, by expecting a quantum efficiency change via more excitations. Lastly, sensitivity is a function of field strength $S(E) = \frac{c_c(E)}{c_c(E)+c_q(E)}$. In Figure 4, the blue solid curve presents $\Delta V_{max}$ with different initial delay offsets and the red solid curve represents $\Delta V_{min}$. These delay offsets help locate the maximum tunable range, and the result shows that the zero-balanced position has a ~100 pV working range for defining sensitivity to ensure a significant $S(E)$ amplitude for low quantum efficiency. The dashed lines in Figure 4 represent two BOC outputs for different pump levels and amplitude coefficients can be determined via two boundaries. By quantify $S(E)$ TCT can differentiate between perturbative and non-perturbative regimes by analyzing the timing and characteristics of the emitted harmonics.

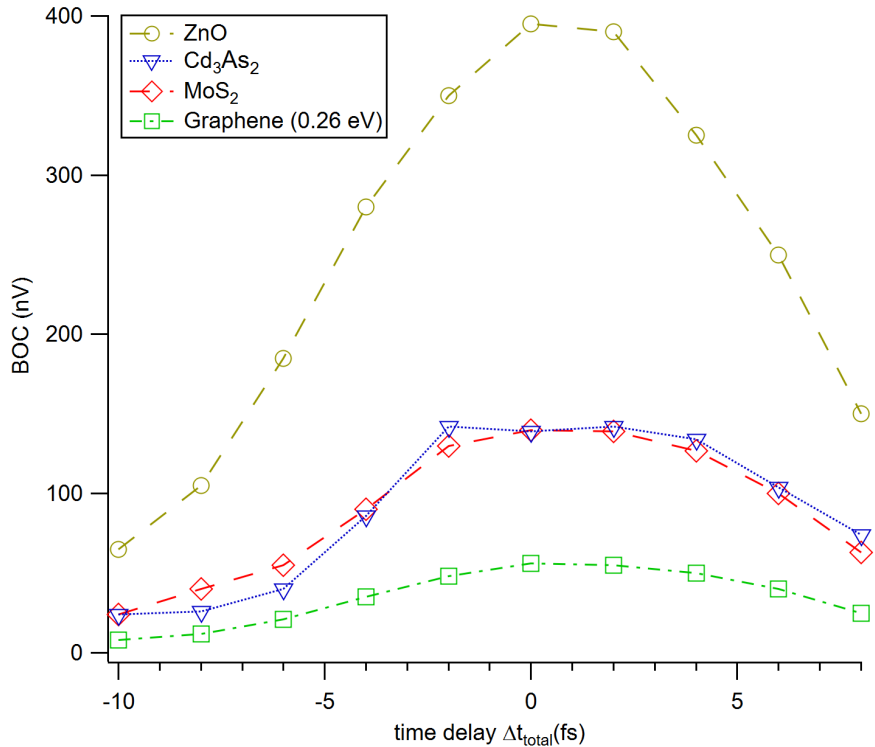

Figure 3: Represents TCT responsivities (nV/fs) of $Cd_3As_2$, $MoS_2$, ZnO, and graphene HHG (at 3rd order of harmonics is closest to the center wavelength 1550 nm) output with respect to the accumulated time delay $\Delta T$ (fs).

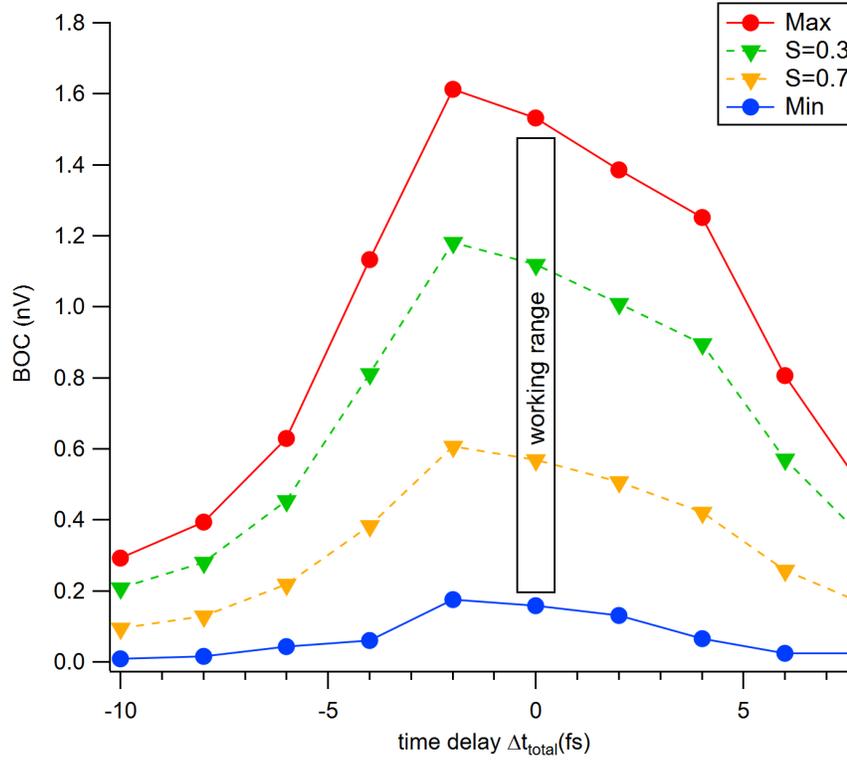

Figure 4: Demonstrates the BOC output (nV) of secondary probing process with respect to the induced time delay $\Delta t_i$ (fs). The solid plots represent two boundaries: the red plot is the BOC output for full-emission that occurs at $\Delta t_r = 10$ fs delay, and the blue plot is the BOC output without the SWP. The dashed line plots simulate BOC output due to partial contribution of stimulated emission; yellow represents $S(E) = 0.3$ and green represents $S(E) = 0.7$. Gray arrow demonstrates a working range (~100 pV) at zero-crossing.

| Materials | Fundamental Wavelength ($\mu m$) | Interested order of Harmonic ($\mu m$) | Zero Crossing (attosecond) | Near balance Sensitivity (nV/fs) |
|---|---|---|---|---|
| $Cd_3As_2$ | 4.5 | 1.50 | 36.15 | 135 |
| $ZnO_2$ | 3.85 | 1.28 | 77.9 | 381 |
| $MoS_2$ | 4.7 | 1.57 | 24.9 | 134 |
| (Mono)graphene | 4.77 | 1.59 | 18 | 47 |

Table I: Parameters of selected quantum materials for HHG in terms of fundamental wavelength and the operating high-harmonic order; zero crossing and near balance sensitivity represent the output of TCT.

## IV. Theory

### *High Harmonic Generation*

To distinguish semi-classical and QED of HHG theoretically we use an SF-QED Hamiltonian:

$$H = 1/2m * (P - qA)^2 + U + H_F. \tag{2}$$

Where, A is defined as the driving laser vector potential, with q being the electron charge, m is the electron mass, U is the atomic potential, and $H_F$ the Hamiltonian of the free electromagnetic field. The quantized vector potential A contains both the driving field and the emitted field, which decomposes to the classical time-dependent part $A_c = <\varphi_{laser}(t)|A|\varphi_{laser}(t)>$ and the additional quantum term $A_q$ the correction term contains all the possible modes of HHG of a single photon transition. To express this additional quantum term we have

$$A_q = \sum_{k\sigma} \sqrt{h/4\pi\varepsilon_0 Vck} \, [e_\sigma a_{\sigma k} e^{ik*r} + e_\sigma^* a_{\sigma k}^\dagger e^{-ik*r}]. \tag{3}$$

Where V is a normalization volume, $\sum_{k\sigma}$ is a summation over all photonic modes with polarization σ and wavevector k, the operators $a_{\sigma k}$ and $a_{\sigma k}^\dagger$ are annihilation and creation operators respectively $e_\sigma$ is a unit vector of polarization and $\varepsilon_0$ is the vacuum permittivity. We have $A_q$ and $A_c$ describe the quantum emitted field and the classical driving field, respectively.

Since the classical vector potential is time-dependent, we can apply the time-dependent Schrodinger equation:

$$ih/2\pi \, \partial|\phi_i(t)> = H_{TDSE}|\phi_i(t)>. \tag{4}$$

Here, $|\phi_i(t)>$ is the wavefunction of the electronic system. Initially $\phi_i(t)$ at $t=-\infty$ typically occupies the ground state or another eigenstate of the electronic system but can also be in a superposition of all eigenstates under strong field conditions, where $A_q$ dominates $A_c$ resulting in a non-perturbed system relates to $c_q$ and $c_c$. Suppose we have an additional weak wave packet at a wavelength equivalent to a particular high-order high harmonic as a secondary probe to disturb an ongoing HHG in the time domain, it can become an indicator of when a strong field condition is reached via whether HHG at the particular order can be shifted temporally through TCT measurement. This could lead to a comprehensive study of quantum materials' critical field strength level, and build a database for further investigation.

*Balanced Optical Cross-Correlation*

The simulation of the BOC employs the fourth-order Runge-Kutta method to model sum frequency generation (SFG) processes, tracking optical beams by their spectral and temporal characteristics as they traverse a simulated type II nonlinear crystal with a PPKPD coefficient for effective quasi-phase-matching. In this setup, optical wave packets, represented by a pair of predefined wave functions, serve as inputs to the SFG process. Through the application of the Runge-Kutta method, a third wave packet emerges as the sum-frequency wave. This process involves simulating a double-pass through the crystal to generate two distinct sum-frequency waves. Owing to chromatic dispersion within the crystal, these sum-frequency waves exhibit variations in intensity, denoted by ΔV, which correspond to the initial temporal discrepancies (Δt) between the two input wave packets, effectively capturing the essence of the BOC technique.

## V. Conclusion

We present an in-depth study on the advancements in high-power ultrafast optical technology and strong-field quantum electrodynamics (SF-QED), particularly focusing on attosecond optical sensing via time correlation transduction (TCT) and high-harmonic generation (HHG). This study highlights the distinction between semi-classical and quantum mechanical descriptions of physical systems, proposing a computational investigation of TCT as a method to capture photon-electron interactions in SF-QED with attosecond precision. TCT allows for the resolution of the dynamic behavior of strong-field phenomena in quantum materials, offering a novel approach for characterizing quantum matter. Demonstrates TCT as a powerful experimental method for attosecond optical sensing, offering insights into the quantum dynamics of strong-field phenomena and paving the way for further exploration of quantum materials and SF-QED physics.

## VI. Acknowledgment

This work was supported in part by the National Science Foundation under Contract No. 2231334, the Department of Energy under Contract Nos. DE-SC0022559 and DE-SC0024096, and the IBM HBCU Quantum Center. Additional support was provided by the Air Force Office of Scientific Research under Contract No. FA9550-23-1-0409 and the Office of Naval Research under Contract No. N00014-24-1-2038. TAS acknowledges support from NSF DMR-2047905. The authors express their gratitude for this essential support, which has significantly contributed to the advancement of this research.